\documentstyle[epsfig,longtable]{aipproc}

\begin{document}

\begin{flushright}	
{PSU-TH/237}\\
{December 2000}\\
\end{flushright}	

\title{NLO parton showers and subtractive techniques\footnote{Talk   
at the Linear Collider Workshop LCWS2000, 
Fermilab, 24-28 October 
2000.}$^{,}$\footnote{Work supported in part by the US 
Department of Energy.}}

\author{F. Hautmann}
\address{Department of Physics, Pennsylvania State University, 
University Park PA 16802}

\maketitle

\begin{abstract}
Motivated by asking how to combine parton showers  
with nonleading QCD matrix elements, we 
discuss a subtractive technique based on gauge-invariant 
Wilson-line operators 
and  how this can be used to treat the soft region.  
\end{abstract}

For precision phenomenology at future high energy colliders, it 
will be valuable to construct Monte Carlo event generators in which 
next-to-leading-order (NLO) QCD corrections can be incorporated in 
parton shower algorithms. There is at present no systematic method 
for doing this. 
Procedures based on 
subtractive methods   have recently been 
proposed \cite{friberg,jccmonte,mcsept}. 
An important step in the implementation of this 
program is to 
show how  to decompose Feynman graphs into sums of terms over
different regions, with the terms  arranged so as to correspond to factors
in a factorization formula  suitable for the Monte Carlo 
application.  

An example of such a decomposition is given in 
 Ref.~\cite{jccmonte}  for 
 the photon-gluon fusion
process in leptoproduction.  
 In this case  soft gluons do 
not enter at the leading power, so that leading regions do not overlap.  
To handle general cases,  however, one needs to treat 
graphs with soft gluons and hence 
with overlapping leading regions. 

Fully numerical~\cite{sopnum} 
or semi-analytical~\cite{catsey} 
subtraction methods have been devised to 
calculate NLO
quantities that are  infrared safe. These methods are not 
directly applicable in   event generators  
that simulate  
the fully exclusive structure  of the hadronic final states, 
 since here the  quantities being computed  are not 
infrared safe in perturbation theory. 
In particular, one cannot use a cancellation of soft gluon 
contributions between real and virtual graphs. 

The technique we discuss in this talk is constructed so that 
the following properties are satisfied~\cite{mcsept}:  
\begin{itemize}
\item[(a)]  The integrand for the hard scattering coefficient 
  is to  be an integrable function,  even when
   the corrections are applied to a process that is not 
  infrared safe.   
\item[(b)] The terms in the expansion of each Feynman graph should arise
  from matrix elements of gauge-invariant operators.
\item[(c)]  In particular, 
  the necessary cut-offs on rapidity integrations should  be
  defined gauge-invariantly.  This involves the use of Wilson lines 
  along non-lightlike directions~\cite{nonlight}.
\item[(d)] The evolution equations~\cite{eveq} with
  respect to these cut-offs should be simple, in the sense that there 
  should be no power-law remainder terms.  
\end{itemize}

In what follows we illustrate this technique using a 
simple example, one-gluon emission graphs in leptoproduction. 

\begin{figure}[b!] 
\centerline{\epsfig{file=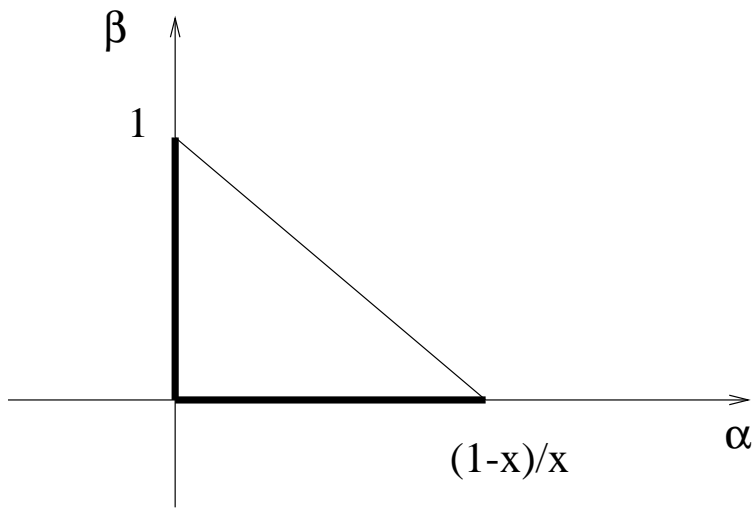
}}
\vspace{10pt}
\caption{The phase space of Eq.~(\ref{rpiece2}) 
in the $\alpha, \beta$ plane. }
\label{fig1}
\end{figure}

We consider a generic 
observable associated with the reaction 
$\gamma^* (q) + q (p) \to g(k) + q(k^\prime)$.   
We denote this  by $\sigma[\varphi]$, where $\varphi$ 
is a weight function 
that contains the definition of the particular observable 
under consideration   
 as a function of the final states
  as well as 
nonperturbative parts of the cross section, including
the parton density.  We denote  by 
 $\alpha$ and $\beta$  the gluon's fractional lightcone momenta,   
$\alpha = k^+ / (x P^+)$, $\beta = k^- /(Q^2/(2 x P^+))$ 
(with $x$ the Bjorken variable, $Q^2$ the photon virtuality, 
$P^+$ the hadron's plus momentum), and by $\phi$ the gluon's    
azimuthal angle. We represent $\sigma[\varphi]$ as    
\begin{equation}
\label{rpiece2}
\sigma [ \varphi] =  
\int_0^\infty d \alpha \int_0^\infty d \beta \ 
\int_0^{2 \pi} {{d \phi} \over {2 \pi}} \,  
 J (x,\alpha , \beta) \,  M (\alpha , \beta) 
\, \varphi (x,Q^2,\alpha , \beta, \phi) \;  
   .
\end{equation} 
Here  
  $ M$ is the next-to-leading-order matrix element. 
For  the 
$\gamma^* q$ 
amplitude 
obtained by contracting the photon Lorentz indices with
the projector corresponding to the structure function $F_2$,  
$ M$ is of the form 
\begin{equation}
\label{matrel2}
 M   (\alpha , \beta) = 
(1 - \beta)^2 \ { { 1 + (1 + \alpha - \beta)^2 } 
\over { \alpha \ 
\beta \ (1 + \alpha - \beta)}} + 2 + 
6 \ {{  (1 - \beta)^2} \over { 1 + \alpha - \beta } }  \; .            
\end{equation} 
The factor $J$ in Eq.~(\ref{rpiece2}) 
is a Jacobian factor \cite{mcsept} and defines the physical 
region in the $\alpha, \beta$ plane as the interior of 
the triangle  in Fig.~1.

The infrared sensitive regions contributing to the 
leading power behavior of 
$\sigma[\varphi]$ are located in Fig.~1 as follows:  
the region in which the gluon is collinear to the initial
state is a neighborhood of the axis $\beta = 0$; the region in which
the gluon is collinear to the final state is a neighborhood of the
axis $\alpha = 0$;  the soft region is a neighborhood of the origin
$\alpha = 0$, $\beta = 0$.  The truly hard region lies away from the
$\alpha = 0$ and $\beta = 0$ axes.

To obtain a
decomposition for $\sigma $ into a sum of terms over the leading regions 
we  employ the method of Ref.~\cite{nonlight}.   
This generalizes the R-operation techniques of renormalization.  
 See  Ref.~\cite{tka} for a related approach.  
We proceed from ``smaller'' to ``larger'' regions. 
For each region $R$, we remove the contribution from smaller regions, 
and construct an approximation to the  matrix element valid in $R$, 
 up to power suppressed corrections. Then we  subtract 
any divergences that appear in this expression coming from 
larger regions. 
To ensure that 
the splitting between the terms is defined gauge-invariantly, 
at each step we demand that the 
counterterms be constructed from matrix elements of Wilson line 
operators, 
\begin{equation}
\label{VIVF}
V_I ( n ) = {\cal P}\exp\left(
 i g \int_{-\infty}^0 dy \, n \cdot A (y \, n ) 
 \right)
\hspace*{0.2 cm} ,  \hspace*{0.3 cm} 
V_{F} ( n ) = {\cal P}\exp\left(
  i g \int^{+\infty}_0 dy \, n \cdot A (y \, n ) 
  \right) ,  
\end{equation}
with suitable directions $n$ for the  lines. 
Evolution equations in  $n$ enable one to   
connect the results corresponding to different 
directions. 
We introduce  light-like directions ${\hat p} = \left( 1, 0, {\bf 0}_T
\right)$, ${\hat p}^\prime = \left( 0, 1, {\bf 0}_T \right)$, and 
non-lightlike directions $u = \left( u^+, u^-, {\bf 0}_T \right)$,
$u^\prime = \left( u^{\prime +}, u^{\prime -}, {\bf 0}_T \right)$ 
(all with  positive components).  
We define $\eta = (2x^2 {P^+}^2/Q^2) u^- / u^+$, and  $\eta' =
(Q^2 / 2x^2 {P^+}^2) {u'}^+ / {u'}^-$. 
 
The result for  the  term  to be associated with the soft 
region reads~\cite{mcsept} 
\begin{equation} 
\label{msoft}
M_S (\alpha, \beta) =  
{2 \over { \alpha  \beta }}  - {2 \over { (\alpha + \eta^\prime 
\beta) \  \beta }} - 
{2 \over { \alpha  \ (\beta + \eta \alpha) }} \; .      
\end{equation}  
The first piece  in the right hand side  is 
just  the soft approximation ($\alpha,\beta \to 0$)  
to Eq.~(\ref{matrel2}). The other two pieces are the subtractions 
associated with the collinear regions where $\alpha/\beta$ or
$\beta/\alpha$ go to zero. 
The first piece 
 can be thought of as the one-loop contribution to the square of a
vacuum--to--gluon matrix element of a product of eikonal Wilson lines
taken along lightlike directions ${\hat p}$, ${\hat
p}^\prime$~\cite{nonlight}.  
 The subtraction pieces can be
derived from operators analogous to those for the first
piece, except for replacing one of the lightlike eikonal lines by a
line along a non-lightlike direction. 
The parameters $\eta$, $\eta^\prime$ 
 act as gauge-invariant 
cut-off parameters. For example, in the second piece, which subtracts the 
divergence from the region $\beta / \alpha \to 0$, $\eta^\prime$ 
provides a cut-off on the region of small $\alpha$.

By applying an analogous treatment we construct terms 
$M_I$, $M_F$ for the 
initial-state and final-state collinear regions. 
The term to be associated with the hard region 
is obtained by subtracting from the original matrix element the 
soft and collinear terms:  
\begin{equation}
\label{subtr}
M_H (\alpha , \beta) = M - M_S - M_I - M_F  \; 
.
\end{equation} 
The 
explicit expression for $M_H$ is given in \cite{mcsept}. This expression 
 is finite in all of the infrared regions,   
and   independent of the choice of the non-lightlike directions 
$u$, $u^\prime$.

In the approach of \cite{jccmonte,mcsept}, $M_H$ gives precisely the 
subtracted hard-scattering function to be used in the Monte Carlo 
to incorporate NLO corrections.  The collinear terms correspond to 
factors in the showering. The soft term can be eliminated by a 
suitable choice of the directions for the Wilson lines: $M_S=0$ for 
 $\eta \eta^\prime = 1$. Then  all 
 of the infrared contributions are 
associated with either 
initial-state or  
final-state branchings, 
 as is appropriate to match the structure of the
parton shower.  

It is interesting to consider the form of the collinear factors 
corresponding to the Monte Carlo scheme for the non-lightlike directions. 
 In the 
 symmetric case $\eta=\eta^\prime = 1$, 
 the result for the initial-state collinear factor 
is~\cite{mcsept}
\begin{equation}
\label{M1}
M_I^{({\rm{MC}})}(\alpha , \beta) =  
  \frac{ 1 }{ \beta } \ 
     \frac{ 1 + (1 + \alpha )^2 }{ \alpha \  (1 + \alpha ) }
  - \frac{ 1 }{ \alpha } \ \frac{ 2 }{ \alpha + \beta } 
\; .   
\end{equation}  
 The first term 
in the right hand side 
corresponds to the standard form of the showering, with 
  the coefficient  of $1/\beta$ given by 
  the  quark $\to$ quark splitting kernel.    
 The second term, on the 
 other hand, is non-standard. However, it  
is suppressed in the collinear region $\beta \to 0$ at 
fixed $\alpha$. That is, this form of the   
 showering   coincides with the 
standard one 
 in the collinear limit  and differs from it away from the 
collinear limit. The role of the second term is to 
 provide an effective cut-off when
 $\alpha \to 0$, i.e., a cut-off on the soft singularity of the 
 splitting kernel.  

Observe  that if one regulated the  $\alpha \to 0$ behavior of 
the first term  in Eq.~(\ref{M1})   
by subtracting its  $\alpha \to 0$ limit, 
given by  $ 2 / ( \alpha  \beta)$, this would bring about an extra  
$\beta \to 0$ singularity. This would not be suited for our  
application in a Monte Carlo algorithm. 
In contrast, the second term in  
Eq.~(\ref{M1}) represents 
precisely what is, from our point of view, 
a  better choice of a counterterm: it subtracts the $\alpha \to 0$ 
singularity without introducing any  extra singular behavior at  
$\beta \to 0$.  

Finally, note that 
this counterterm   cuts off the integration over the region 
of small $\alpha$ at a value of order $\beta$. 
 The procedure 
based on gauge-invariant subtractions   
 tells us precisely  where 
the cut-off is to be placed: the position of the cut-off on $\alpha$ 
turns out to be $\beta$-dependent. In more physical terms, 
this indicates that  the cut-off to be applied in 
the initial-state shower  and the  cut-off to be applied in 
the final-state shower 
are not to be set  independently, but they are related. 

{\bf Acknowledgments.} Thanks to  J.~Collins for 
 collaboration on this topic.

\end{document}